\documentclass{article}

\usepackage{graphicx,epsfig,fullpage,epstopdf,amsmath,amssymb,amsfonts,abstract,layout,indentfirst,amsfonts,cite}
\usepackage[labelfont={bf}, margin=1cm]{caption}
\usepackage[parfill]{parskip}

\newcommand{\ud}{\,\mathrm{d}}
\newcommand{\argmax}{\mathop{\scriptstyle \arg\max}}

\newcommand{\comment}[1]{}

\newcommand{\parsh}[2]{\frac{\partial #1}{\partial #2}}

\newcommand{\nsum}{\sum_{n=1}^N}

\newcommand{\npar}{\eta}
\newcommand{\bnpar}{\boldsymbol{\eta}}
\newcommand{\pre}{\Sigma^{-1}(s)}
\newcommand{\cov}{\Sigma(s)}

\newcommand{\lo}{\lambda_{1}}
\newcommand{\lt}{\lambda_{2}}

\makeatletter
\def\@cite#1#2{[\if@tempswa#2\fi#1]}
\makeatother

\title{Implicit embedding of prior probabilities in optimally efficient neural populations} 
\author{
Deep Ganguli and Eero P. Simoncelli\\[0.7ex]
Howard Hughes Medical Institute, and \\
Center for Neural Science\\
New York University\\
New York, NY 10003 \\[0.7ex]
\texttt{\{dganguli,eero\}@cns.nyu.edu} \\
}

\begin{document}

\maketitle

\begin{abstract}
  \noindent{We examine how the prior probability distribution of a sensory variable in the environment influences the
    optimal allocation of neurons and spikes in a population that represents that variable. We start with a conventional
    response model, in which the spikes of each neuron are drawn from a Poisson distribution with a mean rate governed
    by an associated tuning curve. For this model, we approximate the Fisher information in terms of the density and
    amplitude of the tuning curves, under the assumption that tuning width varies inversely with cell density. We
    consider a family of objective functions based on the expected value, over the sensory prior, of a functional of the
    Fisher information. This family includes lower bounds on mutual information and perceptual discriminability as
    special cases. For all cases, we obtain a closed form expression for the optimum, in which the density and gain of
    the cells in the population are power law functions of the stimulus prior.  Thus, the allocation of these resources
    is uniquely specified by the prior.  Since perceptual discriminability may be expressed directly in terms of the
    Fisher information, it too will be a power law function of the prior.  We show that these results hold for tuning
    curves of arbitrary shape and correlated neuronal variability. This framework thus provides direct and
    experimentally testable predictions regarding the relationship between sensory priors, tuning properties of neural
    representations, and perceptual discriminability.}
\end{abstract}

\section{Introduction} 
Many bottom up theories of neural encoding posit that sensory systems are optimized to represent signals that occur in
the natural environment of an organism \cite{Attneave1954, Barlow1961}.  A precise specification of the optimality of a
sensory representation requires four components: (1) the family of neural transformations (that dictate how natural
signals are encoded in neural activity), over which the optimum is to be taken; (2) the types of signals that are to be
encoded, and their prior distribution in the natural environment; (3) the noise present in the input signals, and the
additional noise that is introduced by the neural transformations; and (4) the costs (e.g., metabolic) of building,
operating, and maintaining the system \cite{Simoncelli2001}.  Although an optimal solution can be computed for some
simple choices of these components (e.g., Linear response models and Gaussian signal and noise distributions
\cite{Atick1990,Doi2012}), the general problem is intractable.

A substantial literature has considered simple population coding models in which each neuron's mean response to a scalar
variable is characterized by a tuning curve \cite[e.g.,~]{Seung1993, Salinas1994, Snippe1996, Sanger1996,Zhang1998,
  Zemel1998, Pouget2003, Jazayeri2006, Ma2006}. For these models, several papers have examined the optimization of
Fisher information, which expresses a bound on the mean squared error of an unbiased estimator \cite{Zhang1999,
  Pouget1999, Brown2006, Montemurro2006}. In these results, the distribution of sensory variables was assumed to be
uniform, and the populations were assumed to be homogeneous with regard to tuning curve shape, spacing, and amplitude.

The distribution of sensory variables encountered in the environment is often non-uniform, and it is thus of interest to
understand how these variations in probability affect the design of optimal populations. It would seem natural that a
neural system should devote more resources to regions of sensory space that occur with higher probability, analogous to
results in coding theory~\cite{Gersho1991}. At the single neuron level, several publications describe solutions in which
monotonic neural response functions allocate greater dynamic range to higher probability stimuli \cite{Laughlin1981,
  Nadal1994, Twer2001, McDonnell2008}. At the population level, non-uniform allocations of neurons with identical tuning
curves have been shown to be optimal for non-uniform stimulus distributions \cite{Brunel1998, Harper2004}.

Here, we examine the influence of a sensory prior on the optimal allocation of neurons and spikes in a population, and
the implications of this optimal allocation for subsequent perception.  Given a prior distribution over a scalar
stimulus parameter, and a resource budget of $N$ neurons with an average of $R$ spikes/sec for the entire population, we
seek the optimal shapes, positions, and amplitudes of the tuning curves. We assume a population with Poisson-like
spiking (which may include correlations), and consider a family of objective functions based on Fisher information. This
family includes lower bounds on mutual information, and the minimum attainable perceptual discrimination performance as
special cases. We then approximate the Fisher information in terms of two continuous resource variables, the density and
gain of the tuning curves. This approximation allows us to obtain a closed form solution for the optimal population. For
all objective functions, we find that the optimal tuning curve properties (cell density, tuning width, and gain) are
power-law functions of the stimulus prior, with exponents dependent on the specific choice of objective
function. Through the Fisher information, we also derive a bound on perceptual discriminability, again in the form a
power-law of the stimulus prior. Thus, our framework provides direct and experimentally testable links between sensory
priors, tuning properties of optimal neural representations, and perceptual discriminability.  This work was initially
presented in ~\cite{Ganguli2010a,Ganguli2010}, and portions appear in the doctoral dissertation of the first author
\cite{Ganguli2012a}.

\section{Encoding model and resource constraints}
We begin with a conventional model for a population of $N$ neurons responding to a single scalar variable, $s$
\cite[e.g.,~]{Seung1993, Salinas1994, Snippe1996, Sanger1996,Zhang1998, Zemel1998, Pouget2003, Jazayeri2006, Ma2006}.
We assume initially that the number of spikes emitted (per unit time) by the $n$th neuron is a sample from an
independent Poisson process, with mean rate determined by its tuning function, $h_n(s)$. The probability density of the
population response can be written as
\begin{align}
  p(\vec{r}|s) = \prod_{n=1}^{N}\frac{h_{n}(s)^{r_n}\ e^{-h_{n}(s)}}{r_{n}!} . \label{pid}
\end{align}
For now, we also assume that the tuning functions can be described by unimodal functions of arbitrary shape.  We
generalize this analysis to the case of monotonic (saturating) tuning curves of arbitrary shape in section
\ref{sec:monotonic}.  And in section \ref{sec:poissonlike}, we consider more general response models that can include
non-Poisson and correlated spiking.

We assume the total expected spike rate, $R$, of the population is fixed, which places a constraint on the tuning
curves:
\begin{align}
\int p(s) \sum_{n = 1}^{N} h_n(s) \ \ud s = R , \label{gainconstrain}
\end{align}
where $p(s)$ is the probability distribution of stimuli in the environment, and can have an arbitrary form. We refer to
this as a sensory prior, in anticipation of its future use in Bayesian decoding of the population response.

\section{Objective function}
We now ask: what is the best way to represent values drawn from $p(s)$ given the limited resources of $N$ neurons and
$R$ total spikes? To formulate a family of objective functions which depend on both $p(s)$, and the tuning curves, we
first rely on Fisher information, $I_f(s)$, which is defined as \cite{Cox1974}
\begin{align*}
  I_f(s) &= - \sum_{\vec{r}} p(\vec{r}|s) \ \frac{\partial^2}{\partial s^2}\log p(\vec{r}|s).
\end{align*}
The Fisher information provides a measure of how accurately the population response represents the stimulus parameter,
based on the encoding model. It has been used to answer theoretical questions about the influence of tuning curve shapes
\cite{Zhang1999, Pouget1999, Shamir2006} and response variability \cite{Abbott1999, Series2004} on the representational
accuracy of population codes. It has also been used in neurophysiological studies to quantify changes in coding accuracy
resulting from changes in tuning curve shapes during adaptation \cite{Dean2005, Durant2007, Gutnisky2008}. For the
independent Poisson noise model, the Fisher information can be expressed analytically as \cite{Seung1993}
\begin{align*}
I_f(s)=\sum_{n=1}^{N}\frac{h_n'^2(s)}{h_n(s)},
\end{align*}
where $h'_n(s)$ is the derivative of the $n^\text{th}$ tuning curve. 

The Fisher information can also be used to express lower bounds on mutual information \cite{Brunel1998}, the variance of
an unbiased estimator \cite{Cox1974}, and perceptual discriminability \cite{Series2009}. Specifically, the mutual
information, $I(\vec{r};s)$, is bounded by:
\begin{align}
I(\vec{r};s) \geq H(s) - \frac{1}{2} \int p(s) \ \log \left(\frac{2\pi
    e}{I_f(s)}\right) \ud s ,  
\label{mi} 
\end{align}
where $H(s)$ is the entropy, or amount of information inherent in $p(s)$, which is independent of the neural
population. The bound is tight in the limit of low noise, which can occur as $N$ increases, $R$ increases, or both
\cite{Brunel1998}.

The Cramer-Rao inequality allows us to express the minimum expected squared stimulus discriminability achievable by any
decoder:
\begin{align}
  \delta^2 \geq \Delta^2\int \frac{p(s)}{I_f(s)} \ud s \label{crBound}.
\end{align}
The constant $\Delta$ determines the performance level at threshold in a discrimination task. The conventional
Cramer-Rao bound expresses the minimum mean squared error of any estimator, and in general requires a correction for the
estimator bias \cite{Cox1974}.  Here, we use it to bound the squared {\em discriminability} of the estimator, as
expressed in the stimulus space.  This has the advantage that it is independent of bias \cite{Series2009}, and that it
is easily (and commonly) measured in perceptual experiments.

We formulate a generalized objective function that includes the Fisher bounds on information and discriminability as
special cases:
\begin{align}
\argmax_{h_n(s)} \int p(s)\  f\left( \sum_{n=1}^{N}\frac{h_n'^2(s)}{h_n(s)} \right) \ud s ,
\qquad\quad \text{s.t.} 
\quad \int p(s) \sum_{n = 1}^{N} h_n(s) \ \ud s =R ,
\label{objfun}
\end{align}
where $f(\cdot)$ is either the logarithm, or a power function. When $f(x) = \log(x)$, optimizing Eq.~(\ref{objfun}) is
equivalent to maximizing the lower bound on mutual information given in Eq.~(\ref{mi}). We refer to this as the {\em
  infomax} objective function.  Otherwise, we assume $f(x) = x^{\alpha}$, for some exponent $\alpha$.  Optimizing
Eq.~(\ref{objfun}) with $\alpha = -1$ is equivalent to minimizing the squared discriminability bound expressed in
Eq.~(\ref{crBound}). We refer to this as the {\em discrimax} objective function.

\section{How to optimize?}
The objective function expressed in Eq.~(\ref{objfun}) is difficult to optimize because it is non-convex.  To facilitate
the optimization, we first parameterize a heterogeneous neural population by warping and rescaling a homogeneous
population, as specified by a {\em cell density} function, $d(s)$, and a {\em gain} function, $g(s)$.  In the resulting
warped population, the tuning widths are inversely proportional to the cell density.  Second, we show that Fisher
information can be closely approximated by a continuous function of density and gain. Finally, re-writing the objective
function and constraints in these terms allows us to obtain closed-form solutions for the optimal tuning curves.

\subsection{Density and gain for a homogeneous population}
\begin{figure}[t!]
	\begin{center}
		{\includegraphics[]{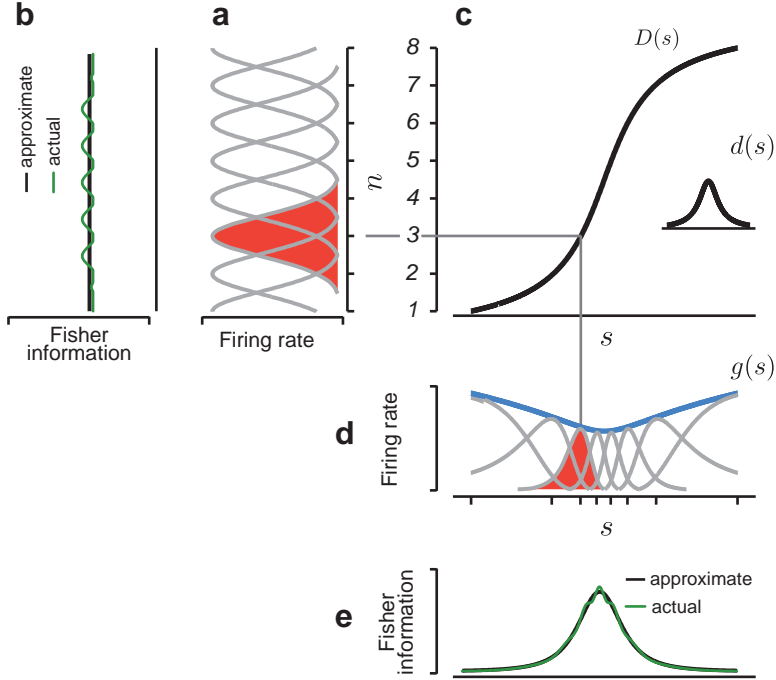}}
        \end{center}
        \caption[Model Parameterization]{Construction of a heterogeneous population of neurons.  {\bf (a)} Homogeneous
          population with Gaussian tuning curves on the unit lattice. The tuning width of $\sigma=0.55$ is chosen so
          that the curves approximately tile the stimulus space.  {\bf (b)} The Fisher information of the convolutional
          population (green) is approximately constant. {\bf (c)} Inset shows $d(s)$, the tuning curve density.  The
          cumulative integral of this density, $D(s)$, alters the positions and widths of the tuning curves in the
          convolutional population.  {\bf (d)} The warped population, with tuning curve peaks (aligned with tick marks,
          at locations $s_n = D^{-1}(n)$), is scaled by the gain function, $g(s)$ (blue). A single tuning curve is
          highlighted (red) to illustrate the effect of the warping and scaling operations. {\bf (e)} The Fisher
          information of the inhomogeneous population is approximately proportional to $d^2(s)g(s)$.}
   \label{fig:optPop}
\end{figure}

If $p(s)$ is uniform, then by symmetry, the Fisher information for an optimal neural population should also be uniform.
We assume a convolutional population of unimodal tuning curves, evenly spaced on the unit lattice, such that they
approximately ``tile'' the space:
\begin{align*}
 \sum_{n = 1}^{N} h(s-n) \approx 1 .
\end{align*}
We also assume that this population has an approximately constant Fisher information:
\begin{align}
I_{f}(s) &= \sum_{n = 1}^{N} \frac{h'^2(s-n)}{h(s-n)} \notag\\
          &= \sum_{n = 1}^{N} \phi(s-n) \approx I_{\rm conv} .
\label{eq:constFisher}
\end{align}
That is, we assume that the Fisher information curves for the individual neurons, $\phi(s-n)$, also tile the stimulus
space.  The value of the constant, $I_{\rm conv}$, is dependent on the details of the tuning curve shape, $h(s)$, which
we leave unspecified.  As an example, Fig.~\ref{fig:optPop}(a-b) shows (through numerical simulation) that the Fisher
information for a convolutional population of Gaussian tuning curves, with appropriate width, is approximately constant.
Now we introduce two variables, a gain ($g$), and a density ($d$), that affect the convolutional population as follows:
\begin{align}
h_n(s) = g\ h\left(d(s-\frac{n}{d})\right)  .
\label{dg}
\end{align}
The gain modulates the maximum average firing rate of each neuron in the population. The density controls both the
spacing and width of the tuning curves: as the density increases, the tuning curves become narrower, and are spaced
closer together so as to maintain their tiling of stimulus space. The effect of these two parameters on Fisher
information is:
\begin{align*}
I_f(s) &= d^2g\sum_{n = 1}^{N(d)} \phi(ds-n)\\
&\approx d^2 g\ I_{\rm conv} .
\end{align*}
The second line follows from the assumption of Eq.~(\ref{eq:constFisher}), that the Fisher information of the
convolutional population is approximately constant with respect to $s$.

The total resources, $N$ and $R$, naturally constrain $d$ and $g$, respectively. If the original (unit-spacing)
convolutional population is supported on the interval $(0,Q)$ of the stimulus space, then the number of neurons in the
modulated population must be $N(d) =Qd$ to cover the same interval.  Under the assumption that the tuning curves tile
the stimulus space, Eq.~(\ref{gainconstrain}) implies that $R=g$ for the modulated population.

\subsection{Density and gain for a heterogeneous population}
Intuitively, if $p(s)$ is non-uniform, the optimal Fisher information should also be non-uniform.  But note that this
could potentially be achieved through inhomogeneities in {\em either} the tuning curve density or gain, and it is not
obvious {\em a priori} what combination of these two functions would yield the best solution.

To solve for an optimal heterogeneous population, we generalize density and gain to be continuous functions of the
stimulus, $d(s)$ and $g(s)$, that warp and scale the convolutional population:
\begin{align}
h_n(s) &= g(s_n)\ h(D(s) -n) .
\label{parampop} 
\end{align}
Here, $D(s) = \int_{-\infty}^{s} d(t) dt$, the cumulative integral of $d(s)$, warps the shape of the prototype tuning
curve. The value $s_n = D^{-1}(n)$ represents the preferred stimulus value of the (warped) $n$th tuning curve
(Fig.~\ref{fig:optPop}(b-d)). Note that the warped population retains the tiling properties of the original
convolutional population.  As in the uniform case, the density controls both the spacing and width of the tuning curves.
This can be seen by rewriting Eq.~(\ref{parampop}) as a first-order Taylor expansion of $D(s)$ around $s_n$:
\begin{align*}
h_n(s) &\approx g(s_n)\ h(d(s_n)(s-s_n)) ,
\end{align*}
which is a generalization of Eq.~(\ref{dg}).

We can now write the Fisher information of the heterogeneous population of neurons in Eq.~(\ref{parampop}) as
\begin{align}
I_f(s) &= \sum_{n=1}^{N} d^2(s)\  g(s_n)\  \phi(D(s)-n) \label{fishproxa} \\
&\approx d^2(s)\ g(s)\ I_{\rm conv}  .
\label{fishprox}
\end{align}
In addition to assuming that the Fisher information is approximately constant (Eq.~(\ref{eq:constFisher})), we have also
assumed that $g(s)$ is smooth relative to the width of $\phi(D(s)-n)$ for all $n$, so that we can approximate $g(s_n)$
as $g(s)$ and remove it from the sum.  The end result is an approximation of Fisher information in terms of the
continuous parameterization of cell density and gain. As earlier, the constant $I_{\rm conv}$ is determined by the
precise shape of the tuning curves.

As in the homogeneous case, the global resource values $N$ and $R$ will place constraints on $d(s)$ and $g(s)$,
respectively.  In particular, we require that $D(\cdot)$ map the entire input space onto the range $[1,N]$.  Thus, for
an input space covering the real line, we require $D(-\infty)=1$ and $D(\infty)=N$ (or equivalently, $\int d(s) \ud s =
N$).  To attain the proper rate, we use the fact that the warped tuning curves sum to unity (before multiplication by
the gain function), along with Eq.~(\ref{gainconstrain}), to obtain the constraint $\int p(s) g(s) \ud s = R$.

\subsection{Objective function and solution for a heterogeneous population}
Approximating Fisher information as proportional to squared density and gain allows us to re-write the objective
function and resource constraints of Eq.~(\ref{objfun}) as
\begin{align} 
\argmax_{d(s),g(s)} \int p(s)\  f\left(d^2(s)\ g(s)\right) \ud s ,
 \quad\quad \text{s.t.}  
\quad \int d(s) \ud s = N ,
\quad \text{and}
\quad \int p(s) g(s) \ud s = R . 
\label{eq:objective}
\end{align}
A closed-form optimum of this objective function is easily determined using calculus of variations. Specifically, one
can compute the gradient of the Lagrangian, set to zero, and solve the resulting system of equations (see Appendix
\ref{sec:infomaxderive}).  Solutions are provided in Table~\ref{solutions} for the infomax, discrimax, and the general
power cases.

\begin{table}
\begin{center}
\scalebox{1}{
\begin{tabular}{l|l|c|l}
  & \bf Infomax & \bf Discrimax &\bf General \\ \hline 
  \hfill Optimized function: &  $f(x) = \log x$ &  $f(x) = -x^{-1}$ &  $f(x) = -x^{\alpha}$, $\alpha < \frac{1}{3}$ \\ \hline 
  \bf{Density (Tuning width)$^{-1}$} \hfill $d(s)$                        &$Np(s)$          &$\propto
  Np^{\frac{1}{2}}(s)$ & $\propto Np^{\frac{\alpha-1}{3\alpha-1}}(s)$\\
  \bf Gain \hfill $g(s)$                             &$R$   &$\propto Rp^{-\frac{1}{2}}(s)$ &$\propto Rp^{\frac{2\alpha}{1-3\alpha}}(s)$ \\
  \bf Fisher information \hfill $I_f(s)$    &$\propto RN^2p^2(s)$        &$\propto RN^2p^{\frac{1}{2}}(s)$ & $\propto RN^2p^{\frac{2}{1-3\alpha}}(s)$\\
  \bf Discriminability bound\ \ \ \hfill $\delta_{\rm min}(s)$   &$\propto p^{-1}(s)$  &$\propto p^{-\frac{1}{4}}(s)$ &$\propto
  p^{\frac{1}{3\alpha-1}}(s)$ \\
\hline
\end{tabular}}
\end{center}
\caption[Closed form solution for optimal neural populations with unimodal tuning curves.]{Optimal heterogeneous
  population properties, for objective functions specified by 
  Eq.~(\ref{eq:objective}).} 
\label{solutions}
\end{table}

\begin{figure}[t!]
	\begin{center}
		{\includegraphics[]{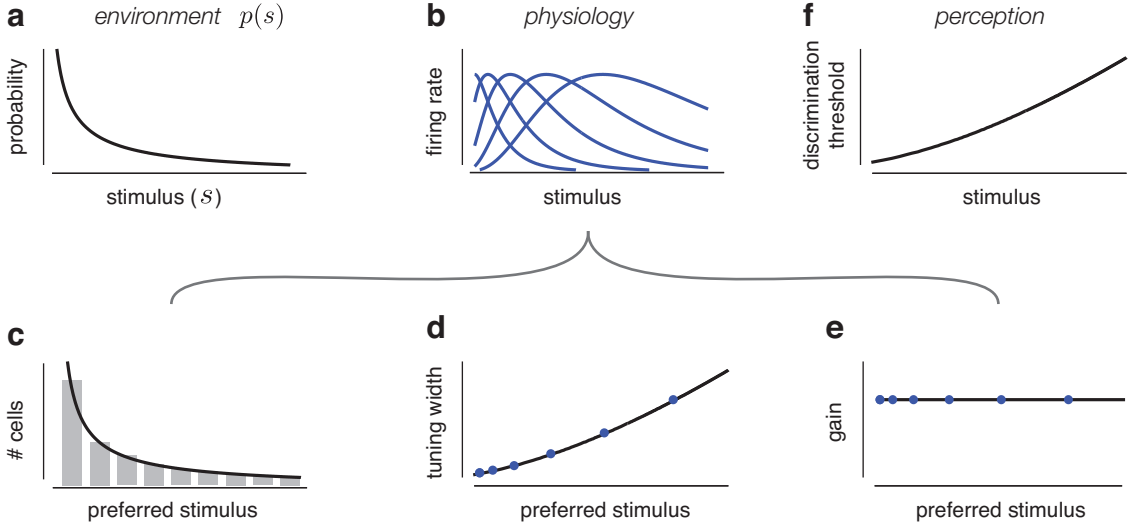}}
        \end{center}
        \caption{ Infomax predictions of the relationship between environment, physiology, and perceptual
          discriminability.  {\bf (a)} An example of a probability distribution over a sensory attribute, $s$, which can
          be directly measured from the environment. {\bf (b)} Tuning curves of a neural population designed to maximize
          the amount of information transmitted about stimuli drawn from the prior distribution in Panel a. {\bf (c-e)}
          Experimentally accessible attributes of the infomax population in Panel b, and their corresponding predictions
          in terms of the prior distribution (black), as expressed in Table 1. {\bf (c)} A histogram of the preferred
          stimuli (stimuli associated with the peaks of the tuning curves) of the neurons is an estimate of local cell
          density, which should be proportional to the prior distribution. {\bf (d)} The tuning widths of the neurons
          (measured as the full width at half maximum of the tuning curves) should be inversely proportional to the
          prior distribution. {\bf (e)} The gain, measured as the maximum average firing rate of each of the neurons,
          should be constant. {\bf (f)} Minimum achievable discrimination thresholds of a perceptual system operating on
          the responses of an infomax population should be inversely proportional to the prior distribution.}
   \label{fig:predictions}
\end{figure}

In all cases, the solution specifies a power-law relationship between the prior, and the density and gain of the tuning
curves. In general, all solutions allocate more neurons, with correspondingly narrower tuning curves, to
higher-probability stimuli. In particular, the infomax solution corresponds to a population with constant gain, and
allocates an approximately equal amount of probability mass to each neuron, as one might intuitively expect from coding
theory (Fig.~\ref{fig:predictions}(a-b)). The shape of the optimal gain function depends on the objective function: for
$\alpha <0$, neurons with lower firing rates are used to represent stimuli with higher probabilities, and for $\alpha
>0$, neurons with higher firing rates are used for stimuli with higher probabilities. Note also that the global resource
values, $N$ and $R$, enter only as scale factors on the overall solution.  As a result, if one or both of these factors
are unknown, the solution still provides a unique specification of the shapes of $d(s)$ and $g(s)$, which can be tested
against experimental data (Fig.~\ref{fig:predictions}(c-e)).

In addition to power-law relationships between tuning properties and sensory priors, our formulation offers a direct
relationship between the sensory prior and perceptual discriminability. This can be obtained by substituting the optimal
solutions for $d(s)$ and $g(s)$ into Eq.~(\ref{fishproxa}), and using the resulting Fisher information to bound the
discriminability, $\delta(s) \geq \delta_{\rm min}(s) \equiv \Delta / \sqrt{ I_f(s)}$ \cite{Series2009}.  The resulting
expressions are provided in Table~\ref{solutions}. In general, the solutions predict that discrimination thresholds
should be lower for more frequently occurring stimuli.  In particular, the infomax solution predicts that inverse
thresholds (discriminability) should be directly proportional to the prior (Fig.~\ref{fig:predictions}(f)). 
\section{Extensions}

\subsection{Monotonic tuning curves} \label{sec:monotonic} 
Thus far we have solved for the optimal cell density and gain for warping and scaling a homogeneous population of
unimodal tuning curves. However, many neurons exhibit \emph{monotonic} tuning to intensity variables such as contrast,
or sound pressure level. The influence of continuous cell density and gain on the Fisher information of a homogeneous
population of monotonic tuning curves is the same as in the unimodal case (Eq.~(\ref{fishprox})), again assuming that
the Fisher information curves of the homogeneous population tile. The constraint on $N$ is also same. However, the total
spiking cost fundamentally differs. Neurons with monotonic tuning curves saturate, and thus the entire population will
be active at the high range of stimulus values, which incurs a large metabolic cost for encoding these
values. Intuitively, this metabolic penalty can be reduced by lowering the gains of neurons tuned to the low end of the
stimulus range, or by adjusting the cell density such that there are more tuning curves tuned to the high end of the
stimulus range. It is not obvious how the reductions in metabolic cost for these coding strategies should trade off with
the optimal coding of sensory information.

To derive the optimal monotonic coding scheme, we first parameterize a heterogeneous population of monotonic tuning
curves by warping and scaling the \emph{derivatives} of a homogeneous population of monotonic tuning curves:
\begin{align}
h_n(s) = \int_{-\infty}^{s}h_n'(t) \ud t =  \int_{-\infty}^{s} g(s_n) d(t) h'(D(t)-n) \ud t. \label{monotonicparms}
\end{align}
This expression is similar to the parameterization of a heterogeneous population of unimodal tuning curves
(Eq.~(\ref{parampop})), except here, $h(\cdot)$ is now a prototype monotonic tuning curve. The density controls both the
number of tuning curves and their slopes, which are inversely proportional to the cell density. The derivatives of the
(warped) monotonic tuning curves, $h'(D(t)-n)$, will be unimodal functions, allowing us to use similar approximations
and intuitions developed for the unimodal case. In particular, we assume that the derivatives of the tuning curves tile
such that $\nsum h'(D(t)-n) \approx 1$.

The total spike count can be expressed from Eqs.~(\ref{gainconstrain} \& \ref{monotonicparms}) as,
\begin{align*}
R &= \int_{-\infty}^{\infty} p(s) \int_{-\infty}^{s} d(t) \nsum g(s_n) h'(D(t)-n) \ud t \ud s. \notag \\
\end{align*}
We define a continuous version of the gain as $g(t) = \nsum g(s_n) h'(D(t)-n)$ which allows us to approximate the total
number of spikes as
\begin{align*}
  R &= \int_{-\infty}^{\infty} p(s) \int_{-\infty}^{s} d(t) g(t) \notag \\
  &= \int_{-\infty}^{\infty} \left(1 - P(s)\right) d(s) g(s) \ud s
\end{align*}
In the second step, we performed integration by parts and defined $P(s) = \int_{-\infty}^{s} p(t) \ud t$ as the
cumulative density function of the sensory prior. The constraint on the total number of spikes is very different than
the bell-shaped tuning curve case, as it now depends on the cell density and the cumulative distribution of the sensory
prior, and will thus affect the optimal solutions for cell density and gain.

We reformulate the original optimization problem of Eq.~(\ref{objfun}) for monotonic tuning curves as:
\begin{align} 
  \argmax_{d(s),g(s)} \int p(s)\ f\left(d^2(s)\ g(s)\right) \ud s , \quad \quad
  &\text{s.t.} \label{eq:objectivemonotonic}
  \quad \hspace{5pt} \int d(s) \ud s = N , \\
  \quad &\text{and} \quad \int \left(1-P(s)\right) d(s) g(s) \ud s = R . \notag
\end{align}
A closed-form optimum of this objective function is easily determined by taking the gradient of the Lagrangian, setting
to zero, and solving the resulting system of equations.  Solutions are provided in Table.~\ref{solutions_monotonic} for
the infomax, discrimax, and general power cases, in addition to solutions for the optimal Fisher information and minimum
achievable discrimination thresholds achievable by a subsequent perceptual system.

\begin{table}
\begin{center}
\scalebox{1}{
\begin{tabular}{l|l|c|l}
  & \bf Infomax & \bf Discrimax &\bf General \\ \hline 
  \hfill Optimized: &  $f(x) = \log x$ &  $f(x) = -x^{-1}$ &  $f(x) = -x^{\alpha}$, $\alpha < \frac{1}{3}$ \\ \hline 
  \bf{Density} \hfill $d(s)$                        &$Np(s)$          &$\propto Np(s)^{\frac{1}{3}}\left[1-P(s)\right]^{\frac{1}{3}}$ & $\propto Np(s)^{\frac{1}{1-2\alpha}}\left[1-P(s)\right]^{\frac{\alpha}{2\alpha-1}}$\\
  \bf Gain \hfill $g(s)$                             &$RN^{-1}\left[1-P(s)\right]^{-1}$   &$RN^{-1}\left[1-P(s)\right]^{-1}$ &$RN^{-1}\left[1-P(s)\right]^{-1}$ \\
  \bf Fisher. \hfill $I_f(s)$    &$\propto RNp^2(s)\left[1-P(s)\right]^{-1}$        &$\propto
  RNp^\frac{2}{3}(s)\left[1-P(s)\right]^{-\frac{1}{3}}$ & $\propto
  RNp^\frac{2}{1-2\alpha}(s)\left[1-P(s)\right]^{\frac{1}{2\alpha-1}}$ \\ 
  \bf Discrim.\ \ \ \hfill $\delta_{\rm min}(s)$   &$\propto p^{-1}(s)\left[1-P(s)\right]^{\frac{1}{2}}$  & $\propto p^{-\frac{1}{3}}(s)\left[1-P(s)\right]^{\frac{1}{6}}$ &$\propto p^{\frac{1}{2\alpha-1}}(s)\left[1-P(s)\right]^{\frac{1}{2-4\alpha}}$ \\
\hline
\end{tabular}}
\end{center}
\caption[Closed form solution for optimal neural populations with monotonic tuning curves.]{Optimal heterogeneous population properties, for objective functions specified by  Eq.~(\ref{eq:objectivemonotonic}).} 
\label{solutions_monotonic}
\end{table}

For all objective functions, the solutions for the optimal density, gain, and discriminability are products of power law
functions of the sensory prior, and its cumulative distribution. In general, all solutions allocate more neurons with
greater dynamic range to more frequently occurring stimuli. Unlike the solutions for unimodal tuning curves
(Table~\ref{solutions}), the optimal gain is the same for all objective functions: for a neuron tuned to a particular
stimulus value, the optimal gain is inversely proportional to the probability of all stimuli occurring after that
stimulus value. Intuitively, this solution allocates lower gains to neurons tuned to the low end of the stimulus range,
which is metabolically less costly. The global resource values $N$ and $R$ again only appear as scale factors in the
overall solution, allowing us to easily compare the predicted relationships to experimental data, even when $N$ and $R$
are not known.

\subsection{Generalization to Poisson-like noise distributions} \label{sec:poissonlike} 
Our results depend on the assumption that neuronal variability is Poisson distributed and neural responses are
statistically independent.  In a Poisson model, the variance of the neural responses is directly proportional to the
mean responses, which has been observed experimentally in some cases \cite{Britten1993}, but may not be true in
general. In addition, the assumption that neuronal responses are statistically independent conditioned on the stimulus
value is often violated \cite{Zohary1994, Kohn2005}.

Here, we generalize the results to a family of ``Poisson-like'' response models \cite{Ma2006,Beck2007}, that allow for
stimulus dependent correlations and an arbitrary linear relationship between mean and variance of the population
response.  We assume the probability density of the population response can be written as
\begin{align} 
  p(\vec{r}|s) &= f(\vec{r})\exp\left[\bnpar(s)^T\vec{r}-A(\bnpar)\right]. \label{eq:poisslike}
\end{align}
This distribution belongs to the exponential family with linear sufficient statistics where the parameter $\bnpar(s)$ is
a matrix of the natural parameters of the distribution with the $n^{\text{th}}$ column equal to $\npar_n(s)$,
$A(\bnpar)$ is a (log) normalizing constant that ensures the distribution integrates to one, and $f(\vec{r})$ is an
arbitrary function of the firing rates. The independent Poisson noise model considered in Eq.~(\ref{pid}) is a member of
this family of distributions with parameters: $\bnpar(s) = \log {\bf h(s)}$ where ${\bf h(s)}$ is a matrix of tuning
curves with the $n^{\text{}th}$ column given $h_n(s)$, $A(\bnpar) = \sum_{n=1}^N \exp(\npar_n)$, and $f(\vec{r}) =
\prod_{n=1}^{N}\frac{1}{r_n!}$.

All of our objective functions depend on an analytical form for the Fisher information in terms of tuning curves, which
is then expressed in terms of density and gain. To derive the Fisher information for the response model in
Eq.~(\ref{eq:poisslike}), we start by noting that the derivative of natural parameters is related to the stimulus
dependent covariance matrix of the population responses, $\cov$, and the derivative of the tuning curves as
\cite{Ma2006,Beck2007},
\begin{align}
\parsh{\bnpar}{s} = \pre\parsh{{\bf h}}{s}. \label{eq:nparderivative}
\end{align}
The term $\pre$ is the inverse of the covariance matrix, and is often referred to as a precision matrix.

The Fisher information \emph{matrix} about the natural parameters is simply equal to the covariance matrix
\cite{Cox1974},
\begin{align}
I_f\left[\bnpar(s)\right] = \cov. \label{eq:nparfish}
\end{align}
The local Fisher information about the stimulus, $s$, can be derived from the chain rule as,
\begin{align*}
I_f(s) &= \parsh{\bnpar}{s}^TI_f\left[\bnpar(s)\right]\parsh{\bnpar}{s}. \\
\end{align*}
After substituting the relationships in Eq.~(\ref{eq:nparderivative} \& \ref{eq:nparfish}) into this expression we
obtain the final expression for the local Fisher information
\begin{align}
I_f(s) &= \parsh{{\bf h}}{s}^T\pre\parsh{{\bf h}}{s}. \label{eq:sfish}
\end{align}
%

The influence of Fisher information on coding accuracy is now directly dependent on knowledge of stimulus dependent
(inverse) covariance matrix. Estimating such a precision matrix from experimental data is technically challenging
(although see \cite{Kohn2005}). Here, we assume a biologically plausible precision matrix that allows for neuronal
variability to be proportional to the mean firing rate, and the responses of nearby neurons to be correlated
\cite{Abbott1999}. For a homogeneous neural population, $h_n(s) = h(s-n)$, we express each element in the precision
matrix as,
\begin{align}
\Sigma_{n,m}^{-1}(s) =  \frac{\alpha\delta_{n,m} + \beta(\delta_{n,m+1} +\delta_{n+1,m})}{\sqrt{h(s-n)h(s-m)}}. \label{eq:pre}
\end{align}
The parameter $\alpha$ controls a linear relationship between the mean response and the variance of the response for all
the neurons. The parameter $\beta$ controls the degree of the correlations, and $\delta_{n,n} = 1$ for all $n$ while
$\delta_{n,m} = 0$ if $n \neq m$. The Fisher information of a homogeneous population may now be expressed from
Eqs.~(\ref{eq:sfish} \&\ref{eq:pre}) as,
\begin{align*}
  I_f(s) &= \alpha \sum_{n=1}^{N} \phi(s-n) \hspace{6pt}+ \beta \hspace{-.25cm}\sum_{n,m = n \pm 1}
  \frac{h'(s-n)h'(s-m)}{\sqrt{h(s-n)h(s-m)}} \\
&\approx \alpha I_\text{conv} + \beta I_\text{corr}
\end{align*}
In the last step we make two assumptions. First, we assume (as for the independent Poisson case) the Fisher information
curves, $\phi(s-n)$, of the homogeneous population tile such that they sum to the constant, $I_\text{conv}$.  Second, we
assume that the cross terms, $\frac{h'(s-n)h'(s-m)}{\sqrt{h(s-n)h(s-m)}}$, also tile such that they sum to the constant,
$I_{\text{corr}}$.

The Fisher information for a heterogeneous population, obtained by warping and scaling the homogeneous population by the
density and gain is
\begin{align} 
  I_f(s) &= d^2(s)\alpha \sum_{n=1}^{N} g(s_n) \phi(D(s)-n)  \\
  &\hspace{12pt}+ d^2(s)\beta \hspace{-.25cm}\sum_{n,m = n \pm
    1} \frac{g(s_n)g(s_m)}{\sqrt{g(s_n)g(s_m)}} \frac{h'(D(s)-n)h'(D(s)-m)}{\sqrt{h(D(s)-n)h(D(s)-m)}}  \notag\\
  &\approx d^2(s)g(s)\left[\alpha I_\text{conv}+\beta I_\text{corr}\right]. \label{eq:heterofishcorr}
\end{align}
In the second step we make three assumptions. First, (as for the independent Poisson case) we assume $g(s)$ is smooth
relative to the width of $\phi(D(s)-n)$ for all $n$, so that we can approximate $g(s_n)$ as $g(s)$. Second, we assume
that the neurons are sufficiently dense such that $\frac{g(s_n)g(s_m)}{\sqrt{g(s_n)g(s_m)}} \approx g(s_n)$. Finally, we
assume $g(s)$ is also smooth relative to the width of the cross terms. As a result, the gain factors can be approximated
by same the continuous gain function, $g(s)$, and can be pulled out of both sums.

Given the form of the Fisher information (Eq.~(\ref{eq:heterofishcorr})), we conclude that the optimal solutions for the
density and gain are the same as those expressed in Tables~\ref{solutions_monotonic} \& \ref{solutions}, which were
derived for an independent Poisson noise model ($\alpha = 1, \beta = 0$). The values of the Fisher information, and
minimum achievable discrimination thresholds now depend on three additional scale factors, $\alpha$, $\beta$, and
$I_\text{corr}$, that characterize the correlated variability of the population code.

\section{Discussion}
We have examined the influence of sensory priors on the optimal allocation of neural resources, as well as the
implications of this optimal allocation on subsequent perception. For a family of objective functions, we obtain
closed-form solutions specifying power law relationships between the prior probability distribution of a variable in the
environment, the tuning properties of a population that encodes that variable, and the minimum perceptual discrimination
thresholds achievable for that variable.  The predictions are easily testable, and preliminary evidence indicates that
the infomax solution is consistent with physiological and perceptual data for several sensory attributes
\cite{Ganguli2010,Ganguli2012a}.

Our analysis requires several approximations and assumptions in order to arrive at an analytical solution. First, we
rely on lower bounds on mutual information and discriminability, each based on Fisher information. Fisher information is
known to provide a poor bound on mutual information when there are a small number of neurons, a short decoding time, or
non-smooth tuning curves \cite{Brunel1998,Bethge2002}. It also provides a poor bound on supra-threshold discriminability
\cite{Shamir2006,Berens2009}.  It is worth noting, however, we do not require the bounds on either information or
discriminability to be tight, but rather that their optima be close to that of their corresponding true objective
functions.  In addition, our preliminary evidence indicates that, at least for typical experimental settings, both
physiological and perceptual data appear to be consistent with the infomax version of our theory
\cite{Ganguli2010,Ganguli2012a}.  We also made several assumptions in deriving our results: (1) the tuning curves,
$h(D(s)-n)$, or in the monotonic case their derivatives, $h'(D(s)-n)$, evenly tile the stimulus space; (2) the single
neuron Fisher informations, $\phi(D(s)-n)$, evenly tile the stimulus space; and (3) the gain function, $g(s)$, varies
slowly and smoothly over the width of $\phi(D(s)-n)$. These assumptions allow us to approximate Fisher information in
terms of cell density and gain (Fig. ~\ref{fig:optPop}(e)), to express the resource constraints in simple form, and to
obtain a closed-form solution to the optimization problem.

Our framework offers an important generalization of the population coding literature, allowing for non-uniformity of
sensory priors, and corresponding heterogeneity in tuning and gain properties. Nevertheless, it suffers from the main
simplification found throughout that literature: the tuning curve response model is restricted to a single
(one-dimensional) stimulus attribute.  Real sensory neurons exhibit selectivity for multiple attributes.  If the
environmental distribution (prior) for those attributes is separable (i.e., if the values of those attributes are
statistically independent) then an efficient code can be constructed separably.  That is, each neuron could have joint
tuning arising from the product of a tuning curve for each attribute. But extending the theory to handled multiple
attributes with statistical dependencies is not straightforward.

Our formulation assumes the sensory attribute of interest is drawn from a fixed and stable distribution.  However, the
distribution of sensory inputs can vary according to context, and it is of interest to consider how the theory might be
generalized to adjust to such changes.  A potential clue comes from the physiology: a large body of literature describes
adaptive changes in neural gain at time scales ranging from milliseconds to hours.  These have been interpreted as
homeostatic mechanisms whose purpose is to maintain a high level of information
transmission~\cite{Wainwright1999,Brenner2000,Fairhall2001}.  How does this fit with our predictions?  A potential
interpretation is that tuning curves, which presumably arise from the strength of synaptic connections, are established
and adjusted over slow time scales, so as to efficiently capture the heterogeneities in stable environmental
distributions, whereas the gains of individual neurons are adjusted more rapidly, so as to adapt to fluctuating
heterogeneities in input intensity, local metabolic resources, or task-related requirements.

Finally, the structure of our optimal population has direct implications for Bayesian decoding, a problem that has
received much attention in recent literature \cite[e.g.,~] {Knill2004,Ma2006,Jazayeri2006,Simoncelli2009,Fiser2010}.  A
Bayesian decoder must have knowledge of prior probabilities, but an often-overlooked issue is how such knowledge is
obtained and represented in the brain \cite{Simoncelli2009}.  Our efficient coding solution provides a mechanism whereby
the prior is implicitly encoded in the arrangement and gains of tuning curves.  Recent publications
\cite{Shi2009,Fischer2011,Girshick2011,Wei2012} have proposed that a population-vector computation (i.e., the average of
the preferred stimuli of the neurons, weighted by their corresponding responses), coupled with an inhomogeneous
arrangement of tuning curves, could provide a simple means for the brain to approximate a Bayesian estimate.  For the
case of an infomax population, we have derived a decoder that more closely approximates a Bayesian least-squares
estimator. Similar to the population vector, it computes a weighted average of the preferred stimuli, but the weights
are constructed by linearly combining and then exponentiating the responses \cite{Ganguli2012,Ganguli2012a}.  Thus,
efficient population representations may offer unforeseen benefits for explaining subsequent stages of sensory
processing.
 
\appendix
\section{Solution for the infomax objective function with bell-shaped tuning curves}
\label{sec:infomaxderive}
The optimum of the objective function in Eq.~(\ref{eq:objective}) is easily determined using the method of Lagrange
multipliers. As an example, consider the case when $f(\cdot) = \log(\cdot)$, which corresponds to optimization of the
Fisher bound on mutual information between the input signal and the population response.  The Lagrangian for this case
is expressed as:
\begin{align*}
L(d(s), g(s), \lo, \lt) = \int p(s) \log{\left( d^2(s) g(s) \right)} \ud s + \lo\left(\int d(s) \ud s - N\right) +\lt\left(\int p(s) g(s)
\ud s - R\right).
\end{align*}
The optimal cell density and gain that satisfy the resource constraints are determined by setting the variational
gradient of the Lagrangian to zero, and solving the resulting system of equations:
\begin{align}
\parsh{L}{d(s)} &= 2p(s)d^{-1}(s) + \lo  = 0 \label{eq:solved}\\
\parsh{L}{g(s)} &= p(s)g^{-1}(s) + \lt p(s) = 0  \label{eq:solveg}\\
\parsh{L}{\lo} &= \int d(s) \ud s - N = 0 \label{eq:lambda1}\\
\parsh{L}{\lt} &= \int p(s) g(s) - R = 0. \label{eq:lambda2}
\end{align}
The optimal cell density and gain are determined from Eqs.~(\ref{eq:solved} \& \ref{eq:solveg}) as:
\begin{align}
d(s) &= -2\lo^{-1} p(s)  \label{eq:oden}\\
g(s) &= -1\lt^{-1} \label{eq:ogain}
\end{align}
The unknown Lagrange multipliers can be determined by substituting these expressions into Eqs.~(\ref{eq:lambda1} \&
\ref{eq:lambda2}) and solving. The result is:
\begin{align*}
\lo &= -2N^{-1} \\
\lt &= -R^{-1}
\end{align*}
Substituting these expressions into Eqs.~(\ref{eq:oden} \& \ref{eq:ogain}) yields the Infomax solution expressed in
Table~\ref{solutions}. The same method was used to derive the rest of the solutions expressed in Tables~\ref{solutions} \& \ref{solutions_monotonic}.

\bibliography{/Users/dganguli/Documents/zotero}
\bibliographystyle{unsrt}

\end{document}